\documentclass[letterpaper, 10 pt, journal, twoside]{ieeetran}  

\IEEEoverridecommandlockouts                              

\usepackage[usenames,dvipsnames]{color}
\usepackage{amsmath,amsfonts}
\usepackage{bbm}
\usepackage{amsmath,bm}
\usepackage{amssymb}
\usepackage{accents}
\usepackage{cite}

\newcommand{\alias}{C:/Users/yishi/Dropbox/bib/alias}
\newcommand{\New}{C:/Users/yishi/Dropbox/bib/New}
\newcommand{\Main}{C:/Users/yishi/Dropbox/bib/Main}
\newcommand{\FP}{C:/Users/yishi/Dropbox/bib/FP}

\usepackage{tikz-timing}
\usepackage{tikz}
\usepackage{subfigure}
\usepackage[marginal]{footmisc}
\usetikzlibrary{arrows,positioning,automata}

\newcommand{\RS}{\mathbb{S}}

\newcommand{\RomanNumeralCaps}[1]
{\MakeUppercase{\romannumeral #1}}

\newtheorem{theorem}{Theorem}
\newtheorem{definition}{Definition}
\newtheorem{lemma}{Lemma}
\newtheorem{assumption}{Assumption}
\newtheorem{proposition}{Proposition}

\newcommand\SmallMatrix[1]{{%
		\scriptsize\arraycolsep=0.4\arraycolsep\ensuremath{\begin{bmatrix}#1\end{bmatrix}}}}

\newcommand\SmallMatrixNBla[1]{{%
		\small\arraycolsep=0.9\arraycolsep\ensuremath{\begin{matrix}#1\end{matrix}}}}

\usepackage{graphicx}
\usepackage{epstopdf}

\usepackage{mathtools}

\newcommand{\R}{\mathbb{R}}
\newcommand{\BL}{\color{black}}

\makeatletter
\let\NAT@parse\undefined
\makeatother
\usepackage{hyperref}  
\hypersetup{colorlinks,linkcolor={blue},citecolor={red},urlcolor={red}} 
\title{\LARGE \bf
	Mediated Remote Synchronization of Kuramoto-Sakaguchi Oscillators: the Number of Mediators Matters
}

\author{Yuzhen Qin, Ming Cao, Brian D.O. Anderson, Danielle S. Bassett, and Fabio Pasqualetti
	\thanks{ This work was supported in part by awards ARO W911NF-18-1-0213, ARO 71603NSYIP, and NSF NCS-FO-1926829. B.D.O. Anderson was supported by the Australian Research Council under grants DP160104500 and DP190100887, and Data61-CSIRO. Y. Qin and F. Pasqualetti are with the Department of Mechanical Engineering, University of California at Riverside, CA, USA  (\{yuzhenqin, fabiopas\}@engr.ucr.edu). M. Cao is with Engineering and Technology Institute (ENTEG), University of Groningen, the Netherlands (m.cao@rug.nl). B.D.O. Anderson is with School of Automation, Hangzhou Dianzi University, Hangzhou, China, and Data61-CSIRO and {Research School of Electrical, Energy and Materials Engineering,} Australian National University, Canberra, Australia (brian.anderson@anu.edu.au). D. S. Bassett is with the Department of Bioengineering, the Department of Electrical and Systems Engineering, the Department of Physics and Astronomy, the Department of Psychiatry, and the Department of Neurology, University of Pennsylvania, PA, USA, and Santa Fe Institute, Santa Fe, NM, USA (dsb@seas.upenn.edu).   
	}	
}

\begin{document}

\maketitle
\thispagestyle{empty}
\pagestyle{empty}

\begin{abstract}
Cortical regions without direct neuronal connections have been observed to exhibit synchronized dynamics. A recent empirical  study has further revealed that such regions that share more common neighbors are more likely to behave coherently. To analytically investigate the underlying mechanisms, we consider that a set of $n$ oscillators, which have no direct connections, are linked through $m$ intermediate oscillators (called \textit{mediators}), forming a complete bipartite network structure.  Modeling the oscillators by the Kuramoto-Sakaguchi model, we rigorously prove that  \textit{mediated remote synchronization}, i.e., synchronization  between those $n$ oscillators that are not directly connected, becomes more robust as the number of mediators increases. Simulations are also carried out to show that our theoretical findings can be applied to other general and complex networks. 
\end{abstract}


\section{INTRODUCTION}

\IEEEPARstart{S}{ynchronization} has been pervasively observed in the human brain. Synchronized central pattern generates (CPGs)  drive coordinated locomotion behaviors \cite{AJI:08}. Particularly, synchrony between cortical regions is believed to facilitate neuronal communication \cite{PJ-PS-KK:05}. Various patterns of synchronization have been observed for  different cognitive tasks that require distinct communication structure \cite{FP:05}.  Also, transient patterns of synchrony can subserve information routing between cortical regions \cite{PA-GT-WF-BD:17}. The underlying anatomical brain network has been shown to play a fundamental role in shaping various patterns of synchrony \cite{CJH-RK-MB-OS:07}. 
 Interestingly, there exists strong evidence that cortical regions without direct axonal links exhibit  synchrony \cite{VF-LJ-RE-MJ:01}. Such synchronization is known as \textit{remote synchronization}. Morphological symmetry in the anatomical network is a mechanism besides some others, e.g., cytoarchitectonic similarity \cite{MY-OT-SH:16} and gene co-expression \cite{BRF-MJD-KAE-SJ-SDR-BDS:19}, that are  believed to account for the emergence of remote synchronization \cite{NV-VM:13}.

 It is shown in \cite{RV-LLG-CRM-IF-GP:08} that two distant neuronal regions symmetrically connected through a third one surprisingly display zero-lag synchronization even in the presence of large synaptic delays. The third region, acting as a mediator (a term used in \cite{LZ-AEM-TN:17}), plays a crucial role. A recent empirical study further shows that the level of synchrony between two remote regions  significantly correlates with the number of such mediators in the anatomical network \cite{VV-PH:14}. However, a theoretical explanation is still missing, which motivates us to analytically investigate the effect of the number of mediators on remote synchronization. With this aim, we single out the mediator-mediated structure from complicated brain networks and consider a simplified and analytically tractable type of network (i.e., a complete bipartite network with two disjoint sets of size $n$ and $m$, respectively). We then study how stable remote synchronization can arise between the set of $n$ oscillators through the mediation of the other oscillators set (which we refer to as \textit{mediators}).

\textbf{Related work}: While complete synchronization has been extensively studied (see \cite{DF-BF:14} for a survey), some attention has recently also been paid to partial or cluster synchronization due to its broad applications \cite{TM-GB-DSB-FP:19a,Pecora2014,SF-PLM-HAM-MTE-RR:16,QY-KY-PO-CM:19}. As a particular form of partial synchronization, remote synchronization has also attracted considerable interest, e.g., \cite{BA-MS-GB-AN-EJ-FL-KJ:12,GL-CA-FA-FL-GJ-FM:13,YQ-YK-MC:18b}. Particularly, some studies are dedicated to remote synchronization in bipartite networks or networks with bipartite subgraphs, in which the effects of time delays \cite{PN-US-AFM-RR:15} and parameter mismatch of mediators \cite{GLA-FMF-BS:16} are investigated. Yet the influence of the number of mediators remains unknown. To analytically study this influence, we employ the Kuramoto-Sakaguchi model \cite{SH-KY:86} to describe cortical oscillations.  {\BL Unlike amplitude-phase models such as the Stuart-Landau model, the Kuramoto-Sakaguchi does not model amplitude dynamics. Therefore, it can only be used in some circumstances where amplitudes of cortical oscillations are ignored for simplification. In \cite{GL-CA-FA-FL-GJ-FM:13}, amplitudes are believe to be crucial in giving rise to stable remote synchronization. However, time delays are not considered in that study. By contrast, time delays are taken into account in the Kuramoto-Sakaguchi model since the phase shift term is often used to model small synaptic delays \cite{HFC-IEM:12},\cite{PMJ-ADM:15}.} Numerical studies show that the Kuramoto-Sakaguchi model, although it ignores amplitude dynamics,  can reproduce remote synchronization of brain regions observed in empirical data \cite{NV-VM:13,VV-PH:14}. We believe time delays play an important role. 

\textbf{Contributions}: The contribution of this paper is fourfold. First, it is the first attempt, to the best of our knowledge, to theoretically study remote synchronization of Kuramoto-Sakaguchi oscillators coupled by a complete bipartite network. Second, we show that the stability of remote synchronization depends crucially on the phase shift, for which a threshold is identified. A phase shift beyond this threshold can prevent stable remote synchronization. Moreover, this threshold increases with the number of mediators, indicating that more mediators make remote synchronization more robust against phase shifts.  This observation provides an analytical explanation for the simulated and empirical findings in \cite{VV-PH:14}, and help to understand the role of the anatomical network in shaping patterns of synchrony in the brain. Third, in sharp contrast to most of the existing results, e.g., \cite{SJ-FB:18,FD-MC-FB:13short}, which only provide sufficient conditions for the existence of exponentially stable frequency synchronization, we present an almost sufficient and necessary condition. Fourth and finally, we find through simulations that remote synchronization remains stable for any phase shift if there are more mediators than mediated oscillators. Also, our simulation results show that bipartite structure in more complex networks play important roles in facilitating robust remote synchronization.

\textbf{Paper organization}: The remainder of this paper is organized as follows. The considered problems are formulated in Section \ref{ProForm}. Our main results are provided in Section \ref{Main:results}. Some simulation studies are presented in Section \ref{simulation}. Finally, concluding remarks are offered in Section \ref{conclusion}.

\textbf{Notation}: Let $\R$, $\R^{+}$, and $\mathbb N$  denote the sets of reals, positive reals, and positive integers, respectively. Given any $m\in \mathbb N$, let $\mathcal N_m=\{1,2,\dots,m\}$, and let $\mathbf 1_m, \mathbf 0_m$ and $I_m$ denote the $m$-dimensional all-one vector, all-zero vector, and identity matrix, respectively. Let the unit circle be denoted by $\RS^1$, a point of which is phase. Let $\RS^m$ denote the $m$-torus.

\section{Problem Formulation}\label{ProForm}

Consider a network of $N$ coupled oscillators whose dynamics are described by 
	\begin{align}\label{origi}
	\dot \theta_i= \omega_i+\sum\limits_{j=1}^{N}a_{ij}\sin(\theta_j-\theta_i-\alpha),
	\end{align}
where: $\theta_i \in \RS^1$ are the phases of the oscillators; $\omega_i\in \R$ are the natural frequencies; $a_{ij}$ is the coupling strength between oscillators $i$ and $j$; and $\alpha\in (0,\pi/2)$ is the phase shift, which is used to model small synaptic delays \cite{HFC-IEM:12}. Let the graph $\mathcal G=\{\mathcal{V},A\}$ describe the network structure, where $\mathcal V=\{1,\dots, N\}$ is the collection of the nodes, and the weighted adjacency matrix $A=[a_{ij}]$ describes the edges and their weights (there is an edge of weight $a_{ij}$ between oscillators $i$ and $j$ if $a_{ij}>0$). 
In the presence of  $\alpha$, complete synchronization is usually not possible. However, it has been shown that oscillators located at morphologically symmetric positions in a network, despite not being directly connected, can be synchronized. This phenomenon is called \textit{remote synchronization} \cite{BA-MS-GB-AN-EJ-FL-KJ:12}. If the phase shift $\alpha$ is small, then remote synchronization appears to be stable; otherwise, it becomes unstable \cite{NV-VM:13}. 

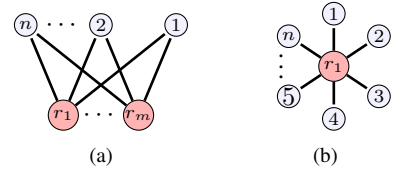
\begin{figure}[!t]
	\centering
	\subfigure[]{\begin{tikzpicture} [->,>=stealth',shorten >=1pt,auto,node distance=0.5cm,
		main node/.style={circle,fill=blue!5,draw,minimum size=0.3cm,inner sep=0pt]},
		red node/.style={circle,fill=red!30,draw,minimum size=0.4 cm,inner sep=0pt]}]
		
		\node[red node]          (0) at (-0.5,0)                      {\scriptsize$r_1$};
		
		\node[red node]          (7) at (0.5,0)                      {\scriptsize$r_{m}$};
		
		\node[main node]          (1) at (1,1.2)         {\scriptsize$1$};		
		\node[main node]          (2) at (0,1.2)     {\scriptsize$2$};	
		\node[main node]          (3) at (-1,1.2)          {\scriptsize$n$};
		

		\tikzset{direct/.style={-,line width=1pt}}
		\path (0)     edge[direct]     node   {} (1)	 
		(0)     edge[direct]     node   {} (2)  
		(0)     edge[direct]     node   {} (3)  
		(7)     edge[direct]     node   {} (1)
		(7)     edge[direct]     node   {} (2)
		(7)     edge[direct]     node   {} (3);			
		\node[] () at(-0.5, 1.2) {$ \cdots $};	
		\node[] () at(0, 0) {$ \cdots $};	
		\end{tikzpicture} 	\label{m_mediator}	}	$\;\;\;\;\;\;$
	\subfigure[]{
		\begin{tikzpicture} [->,>=stealth',shorten >=1pt,auto,node distance=0.5cm,
		main node/.style={circle,fill=blue!5,draw,minimum size=0.3cm,inner sep=0pt]},
		red node/.style={circle,fill=red!30,draw,minimum size=0.4cm,inner sep=0pt]}]
		
		\node[red node]          (0) at (0,0)                      {\scriptsize$r_1$};

		\node[main node]          (1) at (0,0.7)         {\scriptsize$1$};		
		\node[main node]          (2) at (0.6,0.4)        {\scriptsize$2$};	
		\node[main node]          (3) at (0.6,-0.4)          {\scriptsize$3$};
		\node[main node]          (4) at (0,-0.7)          {\scriptsize$4$};	
		\node[main node]          (5) at (-0.6,-0.4)         {$5$};
		\node[main node]          (6) at (-0.6,0.4)         {\scriptsize$n$};

		\tikzset{direct/.style={-,line width=1pt}}
		\path (0)     edge[direct]     node   {} (1)	 
		(0)     edge[direct]     node   {} (2)  
		(0)     edge[direct]     node   {} (3)  
		(0)     edge[direct]     node   {} (5)
		(0)     edge[direct]     node   {} (6)
		(0)     edge[direct]     node   {} (4);			
		\node[] () at(-0.7, 0.1) {$ \vdots $};	
		\end{tikzpicture}		\label{1_mediator}}	
	\caption{Two networks: (a) multiple mediators; (b) one mediator.}
\end{figure}

In this paper we let $\mathcal G$ be a complete bipartite graph (see Fig. \ref{m_mediator}). The dynamics of the oscillators coupled by the network described by $\mathcal G$ then become
\begin{subequations}\label{main:1}
	\begin{align}
	&\dot \theta_i={ \omega_i} +\sum\limits_{q=1}^{m}a_{ir_{q}}\sin(\theta_{r_q}-\theta_i-\alpha), &i\in \mathcal N_n,\label{main:1:o}\\
	&\dot \theta_{r_p}={\omega_{r_p}} +\sum\limits_{j=1}^{n}a_{jr_{p}} \sin(\theta_j- \theta_{r_p}-\alpha), &p\in \mathcal N_m,\label{main:1:m}
	\end{align}
\end{subequations}
where\footnote{We restrict our analysis to the case where $m<n$ in this paper. Outcomes for the case where $m\ge n$ are shown in simulations in Section \ref{simulation}, and suggest interesting theoretical questions.} $m<n$ and $n+m=N$. The peripheral oscillators, $1,\dots,n$, are connected via some intermediate oscillators (colored red in Fig. \ref{m_mediator}). We call those intermediate oscillators \emph{mediators},  since they are mediating the dynamics of the peripheral oscillators. The peripheral oscillators are called \textit{mediated oscillators}. Following \cite{LZ-AEM-TN:17}, we also refer to the synchronization of mediated oscillators, $1,2,\dots,n$, as \textit{mediated remote synchronization}. 
When there is only $1$ mediator, the network reduces to a star (see Fig. \ref{1_mediator}). A threshold of the phase shift $\alpha$, beyond which mediated remote synchronization becomes unstable, has been obtained in \cite{QY-KY-BDOA-CM:19} for a star network with two mediated oscillators. 

We aim to extend this result to a general case in which there can be more than $2$ mediated oscillators (i.e., $n \ge 2$). Interestingly, we also allow for more than $1$ mediator (i.e., $m \ge 1$) and study how the number of mediators affects the threshold for stability on the phase shift $\alpha$.

Let $\theta=(\theta_1,\dots,\theta_n, \theta_{r_1}, \dots, \theta_{r_m})^ \top$, and for any $i$ and $p$ denote the right-hand sides of \eqref{main:1:o} and \eqref{main:1:m} by $f_i(\theta)$ and $g_p(\theta)$, respectively. Then, \eqref{main:1} can be rewritten as $\dot\theta_i=f_i(\theta),\dot \theta_{r_p}=g_p(\theta)$. For simplicity of analysis, we make the following assumption (which later is partially relaxed).
\begin{assumption}\label{Assm:1}
	Assume  $\omega_i=\omega_{r_p}=\omega$ and $a_{i r_p}=1$, $\forall i,p$.
\end{assumption}


Under this assumption, the mediated oscillators are located at symmetric positions. Notice that our results and analysis in the rest of this note remain unchanged if $a_{i r_p}=a$, for any $a>0$, because this operation would preserve symmetry.

 Next, let us first define the \textit{(mediated) remote synchronization manifold}, denoted by $\mathcal M$. For $\theta \in \RS^N$, define $\mathcal M:=\left\{	\theta\in \RS^{N}: \theta_i=\theta_j, \forall i,j\in \mathcal N_n \right\}$. A solution $\theta(t)$ to \eqref{main:1} is said to be \emph{remotely synchronized} if $\theta(t)\in \mathcal M$  for all $t\ge 0$. Note that the phases $\theta_i(t)$ are not required to equal $\theta_{r_p}(t)$ for all $t\ge 0$ in a remotely synchronized solution. We also say that remote synchronization has taken place if a solution to \eqref{main:1} is remotely synchronized. 

Remote synchronization is categorized into two types, depending on whether the phases are locked. A solution is phase-locked if every pair-wise phase difference involving the mediators is constant, or, equivalently, when all the frequencies are synchronized. In contrast, the mediators are allowed to have different frequencies from the $\dot \theta_i$'s in the case of phase-unlocked remote synchronization. We are exclusively interested in studying phase-locked remote synchronization in this paper, and thus we refer to it just as remote synchronization for brevity. The (phase-locked) remote synchronization manifold is then defined as follows.
	
	\begin{definition} \label{def:phase-locked}
		{\bf (Remote Synchronization Manifold)}
	For $\theta \in \RS^N$, the remote synchronization manifold is defined by $\mathcal M_{L}:=\left\{	\theta\in \mathcal M: f_i(\theta)=g_p(\theta), \forall i\in \mathcal N_n, p\in \mathcal N_m  \right\}.$
	\end{definition}

It then suffices to identify the threshold of $\alpha$ beyond which the manifold $\mathcal M_L$ becomes unstable. 

\section{Main Results}\label{Main:results}

In this section, we provide our main results. The threshold for the phase shift $\alpha$, which ensures stability and depends on the numbers of mediated oscillators and mediators, is presented in the following theorem. 

\begin{theorem}\label{mainTheo}
	{\bf{(Threshold of $\alpha$ for stable remote synchronization)}}
	For the dynamics of oscillators \eqref{main:1}, the following statements hold under Assumption \ref{Assm:1}:\\	
	(i) there exists a  unique \emph{exponentially stable} remote synchronization manifold in $\mathcal M_L$ if\footnote{A bifurcation occurs when  $\alpha=\arctan\sqrt{(n+m)/(n-m)}$, but the question of whether there exists an exponentially stable remote synchronization manifold in $\mathcal M_L$ remains unanswered. }

	\begin{align}\label{thre:less}
	\alpha<\arctan \left( \sqrt{\frac{n+m}{n-m}} \right).
	\end{align}	
	(ii) The remote synchronization manifold $\mathcal M_L$ is \emph{unstable} if \begin{align}\label{thre:greater}
	\alpha>\arctan \left( \sqrt{\frac{n+m}{n-m}} \right).
	\end{align}
\end{theorem}

Theorem \ref{mainTheo} implies that a sufficiently large phase shift $\alpha$ (or a large time delay, equivalently) prevents stable remote synchronization. It can be seen that $\arctan \sqrt{({n+m})/({n-m})}$ is monotonically decreasing with respect to $n$ and monotonically increasing with respect to $m$. Thus, a larger $n$ (i.e., more mediated oscillators) results in a narrower range of phase shifts such that an exponentially stable remote synchronization manifold exists. Also, $\lim_{n\to \infty} \arctan \sqrt{(n+m)/(n-m)}={\pi}/{4}$ for any given $m$, which means that exponentially stable remote synchronization \textit{always  exists} regardless of the number of mediated oscillators as long as $\alpha <{\pi}/{4}$. In contrast, a larger $m$ (i.e., more mediators) creates a wider range of $\alpha$ for a given $n$. In other words,  more mediators make remote synchronization \textit{more robust} against phase shifts. Before providing the proof of Theorem \ref{mainTheo}, we present some interesting intermediate results, which will  be used to construct the proof.

\subsection{Intermediate Results}

Unlike the classic Kuramoto model, e.g., \cite{SJ-FB:18,FD-MC-FB:13short}, the usual linearization method cannot be directly used to construct the proof in our case, since the oscillators' frequencies converge to a value that is distinct from the simple average of the natural frequencies \cite{ME-KJ-BB:04} due to the presence of the phase shift $\alpha$.
To overcome this problem, we define some new variables. Let {$ x_i=\theta_{i+1}-\frac{1}{n}\sum_{j=1}^{n}\theta_j$ for all $i \in \mathcal N_{n-1}$, and $y_p=\theta_{r_p}- \frac{1}{n}\sum_{j=1}^{n}\theta_j$ for all $p \in \mathcal N_m$.}  Following \eqref{main:1}, the time derivatives of these new variables are 
\begin{subequations}\label{Phac-diff}
\begin{align}
\dot x_i &\SmallMatrixNBla{= \sum\limits_{q=1}^{m}\sin(y_q-x_{i}-\alpha) -\frac{1}{n}\sum\limits_{j=1}^{n-1}\sum\limits_{q=1}^{m}\sin(y_q-x_j-\alpha)} \nonumber\\
&\SmallMatrixNBla{-\frac{1}{n}\sum\limits_{q=1}^{m}\sin(y_q+\sum\nolimits_{j=1}^{n-1}x_j-\alpha)}, \label{dyn:x_i}\\
\dot y_p& \SmallMatrixNBla{=\sum\limits_{j=1}^{n-1} \sin(x_j-y_p-\alpha)+\sin(-y_p-\sum\nolimits_{j=1}^{n-1}x_j-\alpha)}\nonumber\\
& \SmallMatrixNBla{-\frac{1}{n}\sum\limits_{j=1}^{n-1}} \SmallMatrixNBla{\sum\limits_{q=1}^{m}\sin(y_q-x_j-\alpha)
-\frac{1}{n}\sum\limits_{q=1}^{m}\sin(y_q+\sum\nolimits_{j=1}^{n-1}x_j-\alpha)},\label{dyn:x_0}
\end{align}
\end{subequations}
where $i \in \mathcal N_{n-1}$ and $p\in \mathcal N_m$. Denote $x:=[x_1,\dots,x_{n-1}]^\top \in \RS^{n-1}$ and $y:=[y_1,\dots,y_m]^\top\in \RS^{m}$. 
From Definition \ref{def:phase-locked}, a solution $\theta(t)$ to \eqref{main:1} is remotely synchronized \textit{if and only if}: 1) $x=0$, and 2) $\dot x=0$ and $\dot y=0$. Any $(x,y)$ satisfying 1) and 2) is an equilibrium of \eqref{Phac-diff}. The following proposition states how the stability of remote synchronization in \eqref{main:1} can be analyzed by studying that of the equilibrium points of \eqref{Phac-diff}. 
\begin{proposition}\label{Proposition:equi} {\bf {(Connections between remote synchronization  in \eqref{main:1} and equilibria of \eqref{Phac-diff})}}
	The equilibria that satisfy $x=0$  of the system \eqref{Phac-diff} in $\RS^{N-1}$ are given by{\footnote{The equilibria given in \eqref{equilbrium:1} and \eqref{equilbrium:2} do no exhaust all the possible equilibria of \eqref{Phac-diff}. Other equilibria that do not satisfy $x=0$ may also exist.}}
	\begin{align}
	& \SmallMatrixNBla{e_1=\left[\mathbf{0}^ \top_{n-1},c(\alpha) \mathbf{1}^\top_{m_1}, (\pi-c(\alpha)-2\alpha) \mathbf{1}^\top_{m_2} \right]^\top,} \label{equilbrium:1} \\
	& \SmallMatrixNBla{e_2= \left[\mathbf{0}^ \top_{n-1},(\pi+c(\alpha)) \mathbf{1}^\top_{m_1}, (-c(\alpha)-2\alpha) \mathbf{1}^\top_{m_2} \right]^\top,}\label{equilbrium:2}
	\end{align}
	with 
	\begin{align}\label{expre:c(a)}
	{c(\alpha)=-\arctan\left({ \frac{(n-m_1)\sin \alpha+m_2\sin 3\alpha}{(n+m_1)\cos \alpha+m_2\cos 3\alpha}} \right),}
	\end{align}
	where {$m_1=0,1,\dots,m$, $m_2=m-m_1$ if $m\ge 2$, and $m_1=1,m_2=0$ if $m=1$.}
	There exists a unique exponentially stable remote synchronization manifold in $\mathcal M_L$ \emph{if and only if} one of these equilibria is stable. Furthermore, $\mathcal M_L$ is unstable \emph{if and only if } all the equilibria in \eqref{equilbrium:1} and \eqref{equilbrium:2} are unstable. 
\end{proposition}

\begin{IEEEproof}
	Substituting $x_i=0$ into the right-hand side of \eqref{dyn:x_i} yields, as expected, $\dot x_i=0$ for any $i\in \mathcal N_{n-1}$. Substituting $x_i=0$ into the right-hand side of \eqref{dyn:x_0} leads to $\dot y_{p}=-n\sin(y_p+\alpha)-\sum_{q=1}^{m}\sin(y_q-\alpha)$ for all $p\in \mathcal N_{m}$.
	Since at remote synchronization all $\dot y_p$ are zero, we have $\sin(y_p+\alpha)= -\frac{1}{n} \sum_{q=1}^{m} \sin(y_q-\alpha)$ for any $p$, which means
	$
	\sin(y_p+\alpha)=\sin(y_q+\alpha)
	$
	for any $p,q\in \mathcal N_m$. Then, for a given pair of $p,q$, either $y_p=y_q$ or $y_p+\alpha=\pi-(y_q+\alpha)$ needs to hold. Consequently, at remote synchronization the elements in $y$ are not necessarily identical, but can be clustered into two groups. Assume that the sizes of these two groups are $m_1$ and $m_2$, respectively, where $0\le m_1 \le m$ and $m_1+m_2=m$. Without loss of generality, let $y_p=y_e^*$ for $p=1,\dots,m_1$, and $y_p= \pi-y_e^*-2\alpha$ for $p=m_1+1,\dots,m$. Substituting these $y_p$'s into the equations $-n\sin(y_p+\alpha) - \sum_{q=1}^{m} \sin(y_q-\alpha)=0$ and solving them we obtain two sets of solutions in $\RS^1$, i.e., 1) $y_e^*=c(\alpha)$, and 2) $y_e^*=\pi+c(\alpha)$, where $c(\alpha)$ is given in \eqref{expre:c(a)}.	Then, \eqref{equilbrium:1} and \eqref{equilbrium:2} follow subsequently. Because all the equilibria in \eqref{equilbrium:1} and \eqref{equilbrium:2} together exhaust all the possible equilibria satisfying $x=0$ of \eqref{Phac-diff} in $\RS^{N-1}$ and each corresponds to a remote synchronization manifold in $\mathcal M_L$,  there exists a unique \textit{exponentially stable} remote synchronization manifold in $\mathcal M_L$ if and only if one of these equilibria is stable, and the remote synchronization manifold $\mathcal M_L$ is unstable if and only if all the equilibria are unstable. 
\end{IEEEproof}

For notational simplicity, let $s_1=(n-m_1)\sin \alpha+m_2\sin 3\alpha$, $s_2=(n+m_1)\cos \alpha+m_2\cos 3\alpha$, and $S=\sqrt{s_1^2+s_2^2}$. Then, it follows from \eqref{expre:c(a)} that 
\begin{align}\label{cos:sin}
&\sin c(\alpha)=-\frac{s_1}{S}, &\cos c(\alpha)=\frac{s_2}{S}.
\end{align}
The stability of the equilibria given in \eqref{equilbrium:1} and \eqref{equilbrium:2} can in the first instance be examined using the Jacobian matrix of \eqref{Phac-diff} evaluated at $x=0$:
\begin{align}
&J(y)=\begin{bmatrix}
R_1(y)&\mathbf 0\\
\mathbf 0&R_2(y)\end{bmatrix},\\
{\rm where}\hspace{35pt}&R_1(y)= -\sum\limits_{q=1}^{m}\cos(y_q-\alpha)I_{n-1},\label{Expres:R1}
\\&R_2(y)=-D(y)-C(y),\label{Expres:Rs}
\end{align}
with $C(y)=\mathbf 1_m [\cos(y_1-\alpha),\dots,\cos(y_m-\alpha)]$ and
\begin{align*}  
D(y)=\SmallMatrix{
	n\cos(y_1+\alpha)&0&\dots&0\\
	0&n\cos(y_2+\alpha)&\dots&0\\
	\vdots&\vdots&\ddots&\vdots\\
	0&0&\dots&n\cos(y_m+\alpha)}.
\end{align*}
Accordingly, we investigate the eigenvalues of $J(y)$ at 
\begin{align}
&y= \big[c(\alpha) \mathbf{1}^\top_{m_1}, (\pi-c(\alpha)-2\alpha) \mathbf{1}^\top_{m_2} \big]^\top:=e^y_1, \label{ey1}\\
&y=\big[(\pi+c(\alpha)) \mathbf{1}^\top_{m_1}, (-c(\alpha)-2\alpha) \mathbf{1}^\top_{m_2} \big]^\top:=e^y_2, \label{ey2}
\end{align}
for all the allowed  pairs of $m_1,m_2$. Since $J(y)$ has a block diagonal form, its eigenvalues are composed of  those of $R_1$ and $R_2$. Using this property, we find that some of the equilibria in \eqref{equilbrium:1} and \eqref{equilbrium:2} are always unstable. 

\begin{proposition}\label{prop:unstable} {\bf{(Unstable equilibria)}}
	Let $m\ge 2$. Then, all the equilibria in \eqref{equilbrium:1} and \eqref{equilbrium:2} are \emph{unstable} for any $\alpha \in (0,\pi/2)$ and any $m_1$ satisfying $0\le m_1 < m$.  
\end{proposition}
 
\begin{IEEEproof}
  We construct the proof by showing that for any $m_1$ satisfying $0 \le m_1 <m$ either $R_1(y)$ or $R_2(y)$ has at least one positive eigenvalue no matter whether they are evaluated at $y= e^y_1$ or at $y= e^y_2$.  
  
First, we consider the case when $y= e^y_1$. For any $i$, it follows that
$C_{ij}=\cos(c(\alpha)-\alpha)$ for $j=1, \dots,m_1$, $C_{ij}=-\cos(c(\alpha)+3\alpha)$ for $j=m_1, \dots, m$, and  $D_{ii}=n\cos (c(\alpha)+\alpha)$ for $i=1, \dots,m_1$, $D_{ii}=-n\cos (c(\alpha)+\alpha)$ for $i=m_1+1,\dots,m$. For notational simplicity, let $a=\cos (c(\alpha)+\alpha)$, $b=\cos(c(\alpha)-\alpha)$, and $c=-\cos(c(\alpha)+3\alpha)$, and then $R_1(y), D$, and $C$ can be rewritten as 
  \begin{align}
  & R_1(y)=-((m-1)b+c)I_{n-1},\label{exp:R_1}\\
  &C=\mathbf 1_m [b \mathbf 1^\top_{m_1}, c \mathbf 1^\top_{m_2}], {\rm and\;}D= \SmallMatrix{  na I_{m_1}&\mathbf 0\\
  \mathbf 0&-na I_{m_2}}.\label{exp:C-D}
  \end{align}
  We then show that $D+C$ has a negative eigenvalue in both the following two cases: a) when $m_1\le m-2$; b) when $m_1=m-1$. We start with the case a). Let $v_1:=[\mathbf 0^\top_{m-1}, -1 ,1]^\top$, and then
  $
  (D+C)v_1=Dv_1+\mathbf 0=-na v_1,
  $
  which means that $-na$ is an eigenvalue of $D+C$. As a consequence, $J(y)$ has at least one positive eigenvalue. We then study the case b), and show that either $R_1(y)$ or $R_2(y)$ has a positive eigenvalue. From \eqref{exp:R_1}, all the eigenvalues of $R_1(y)$ are identical and  equal to $-((m-1)b+c)$. To ensure that all the eigenvalues of $J(y)$ have negative real parts, $((m-1)b+c)>0$ needs to hold. We then prove that even when $((m-1)b+c)>0$ the matrix $R_2(y)$ still has a positive eigenvalue. We prove that fact by showing that  there is $\phi$ such that $v_2:=[\mathbf 1^\top_{m-1}, \phi]^\top$ is the eigenvector of $D+C$ and it is associated with a negative eigenvalue (denoted by $\lambda$). Let $(D+C)v_2=\lambda v_2$, and we obtain
  $
  [na \mathbf 1^\top_{m-1},-na \phi]^\top+((m-1)b+c \phi)\mathbf 1_m=\lambda[ \mathbf 1^\top_{m-1},\phi]^\top,
  $
  from which we  have the following two equations
  \begin{equation}\label{equa:lam:phi}
	na+(m-1)b+c \phi=\lambda, -na \phi+(m-1)b+c \phi=\lambda \phi.
  \end{equation} 
  We then show that there is a pair of solutions $\phi$ and $\lambda$ to the above equations, satisfying $\lambda<0$. 
  Canceling $\phi$ in the above equations we obtain the equation of $\lambda$ as follows
  \begin{equation*}
  \lambda^2-\underbrace{((m-1)b+c)}_{w_1}\lambda-na\underbrace{((m-1)b+na-c)}_{w_2}=0.
  \end{equation*}
  The solutions to the above  quadratic equation are
  $ \lambda_{1}=\frac{w_1+ \sqrt{w_1^2+4naw_2}}{2}$ and $\lambda_{2}=\frac{w_1- \sqrt{w_1^2+4naw_2}}{2}.
  $
  Since $w_2>0$ from Proposition \ref{Poly:>0} in the Appendix and $a>0$ from Lemma \ref{lemma:2} in the Appendix, it is not hard to see that $\lambda_{2}<0$. Substituting $\lambda_{2}$ into \eqref{equa:lam:phi}, one can compute the solution $\phi$,  which means that $\lambda_{2}$ is an  eigenvalue of $D+C$ that is associated with the eigenvector $v_2:=[\mathbf 1^\top_{m-1}, \phi]^\top$.
   Therefore, we have proven that $J(y)$ evaluated at $y=e_1^y$ has at least one positive eigenvalue  for any $m_1<m$.
 
 Finally, following similar steps as above one can prove that $J(y)$ evaluated at $y=e_2^y$ also has at least one positive eigenvalue, which completes the proof.   
 \end{IEEEproof}
   
\subsection{Proof of Theorem \ref{mainTheo}}

 \textit{Proof of Theorem \ref{mainTheo}:} 
 From Proposition \ref{Proposition:equi}, we construct the proof by showing that: {\RomanNumeralCaps 1)} when $m_1<m$ (which implies $m\ge 2$, since by definition $m_1=m$ when $m=1$), all the equilibria in \eqref{equilbrium:1} and \eqref{equilbrium:2} are unstable for any $\alpha$, and {\RomanNumeralCaps 2)} when $m_1=m$ (for any $m\ge 1$), $e_1$ is exponentially stable under \eqref{thre:less} and unstable under \eqref{thre:greater}, and $e_2$  is unstable for any $\alpha$.

  First, the proof of {\RomanNumeralCaps 1)} follows directly from Proposition \ref{prop:unstable}.
 
 Second, when $m_1=m$, $c(\alpha)$ in \eqref{expre:c(a)} becomes $c(\alpha)=-\arctan\left(\frac{n-m}{n+m}\tan \alpha\right)$, and $e_1,e_2$ become $e_1=\big[\mathbf{0}^ \top_{n-1},c(\alpha) \mathbf{1}^\top_{m}\big]^\top$ and $e_2=\big[\mathbf{0}^ \top_{n-1},(\pi-c(\alpha)) \mathbf{1}^\top_{m}\big]^\top$, respectively.  
 We prove  {\RomanNumeralCaps 2)} by showing the following two facts: a) $J(y)$, evaluated at $y=e^y_1=c(\alpha)\mathbf{1}^\top_{m}$, is Hurwitz under \eqref{thre:less}, and has positive eigenvalues under \eqref{thre:greater}; b) $J(y)$, evaluated at $y=e^y_2=(\pi-c(\alpha))\mathbf{1}^\top_{m}$, has positive eigenvalues for any $\alpha$.
 
To prove a), we investigate the eigenvalues of $R_1(y)$ and $R_2(y)$ at $y=c(\alpha)\mathbf 1_m$. It follows from \eqref{Expres:R1} and \eqref{Expres:Rs} that 
\begin{align}
\SmallMatrixNBla{R_1(y)}&\SmallMatrixNBla{= -m\cos(c(\alpha)-\alpha)I_{n-1},}\label{Expres:R1:m1=m}\\
\SmallMatrixNBla{R_2(y)}&= \SmallMatrixNBla{
-n\cos(c(\alpha)+\alpha)I_m-\cos(c(\alpha)-\alpha)\mathbf{1}_m\mathbf{1}^ \top _m.} \label{Expres:R2:m1=m}
\end{align}
All the eigenvalues of $R_1$ are $-m\cos (c(\alpha)-\alpha)$. Moreover,
\begin{align}
\SmallMatrixNBla{\cos(c(\alpha)-\alpha)=\frac{1}{S}\left( (n+m)\cos^2\alpha-(n-m)\sin^2\alpha\right) },\label{Cal:R1}
\end{align}
The right-hand side of \eqref{Cal:R1} is positive (negative, respectively) under \eqref{thre:less} (\eqref{thre:greater}, respectively), which means that the eigenvalues of $R_1$ are all negative (positive, respectively). Turning now to $R_2$, observe that ${\rm rank}(\mathbf{1}_m\mathbf{1}^ \top _m)=1$ and $\mathbf{1}_m\mathbf{1}^ \top _m \cdot \mathbf{1}_m=m \mathbf{1}_m$, and thus the matrix $\mathbf{1}_m\mathbf{1}^ \top _m$ has $m-1$ eigenvalues that equal $0$ and one eigenvalue that is $m$. Consequently, given $\eta_1,\eta_2\in \R$, the matrix $\eta_1 I_m+\eta_2 \mathbf{1}_m\mathbf{1}^ \top _m$ has $m-1$ eigenvalues being $\eta_1$ and one eigenvalue being $\eta_1+m\eta_2$.  Denote the eigenvalues of $R_2(y)$ by $\mu_i$, $i\in \mathcal N_m$; then evidently $\mu_i=-n\cos(c(\alpha)+\alpha)$ for $i=1,\dots,m-1$, and $\mu_m=-n\cos(c(\alpha)+\alpha)-m\cos(c(\alpha)-\alpha)$.
For $s_1$ and $s_2$ given in \eqref{cos:sin}, $s_1=(n-m)\sin \alpha$ and $s_2=(n+m)\cos \alpha$, and then
$
n\cos(c(\alpha)+\alpha)+m\cos(c(\alpha)-\alpha)
=\frac{1}{S}\left((n+m)^2\cos^2\alpha+(n-m)^2 \sin^2\alpha\right),\label{cos:-}
$
which is positive for any $\alpha$. Also, $\cos(c(\alpha)+\alpha)>0$ for any $\alpha$ from Lemma \ref{lemma:2} in the Appendix, and thus the eigenvalues of $R_2$ are negative for any $\alpha$. Overall, $J(y)$, evaluated at $y=e_1$,  is Hurwitz under \eqref{thre:less}, and has positive eigenvalues under \eqref{thre:greater}.

We finally prove b). At $y=(\pi+c(\alpha))\mathbf 1_m$, following similar steps as above, the eigenvalues of $R_2(y)$ are 
$
\mu_1=\dots=\mu_{m-1}=n\cos(c(\alpha)+\alpha),\mu_m=n\cos(c(\alpha)+\alpha)+m\cos(c(\alpha)-\alpha).
$
Then, all the eigenvalues of $R_2(y)$ are positive for any $\alpha$, which subsequently means that $J(y)$ has positive eigenvalues for any $\alpha$. The  proof is complete. {\footnotesize \hfill $\blacksquare$}
\vspace{4pt}


	We have proven in Theorem \ref{mainTheo} that exponentially stable remote synchronization is possible  only when $m_1=m$  (under a certain condition on $\alpha$). Note that $m_1=m$ implies that the mediators have an identical phase. Therefore,  to ensure the exponential stability of phase-locked remote synchronization, the mediators themselves have to be synchronized, but their phases usually differ from the mediated oscillators'.  In the phase-unlocked case, however, the mediators can be \emph{incoherent} while still guaranteeing stable remote synchronization, which will be shown numerically in the next section. Finally, the equilibrium $e_1$, when $m_1=m$, is the only equilibrium which can be exponentially stable, and its stability depends only on $R_1(y)$ in \eqref{Expres:R1:m1=m} as $R_2(y)$ is always Hurwitz.  The calculation in \eqref{Cal:R1} shows that, as $\alpha$ approaches $\arctan \sqrt{(n+m)/(n-m)}$ from below, $R_1(y)\to 0$, so that the system has a progressively smaller degree of stability.
	
	If the oscillators and the coupling strengths in \eqref{main:1} are heterogeneous (Assumption \ref{Assm:1} not satisfied), the mediated oscillators usually cannot be exactly synchronized. However, if there are small positive numbers $\delta_\omega$ and $\delta_a$ such that $\max|\omega_i-\omega_j|<\delta_\omega$ and $\max|a_{ir_p}-a_{jr_q}|<\delta_a$, approximate remote synchronization (phases remain close but not identical) occurs, which can be proven by analyzing a perturbed system of \eqref{Phac-diff} using standard perturbation theory \cite[Chap. 9]{HKK:02}.

\begin{figure}[t!]
	\centering		
	\subfigure[]{
\includegraphics[scale=0.62]{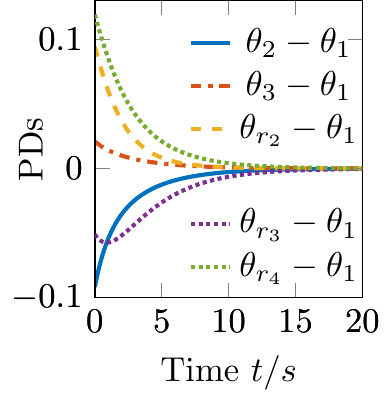}
		\label{Fig:3}}
	\subfigure[]{
		\includegraphics[scale=0.62]{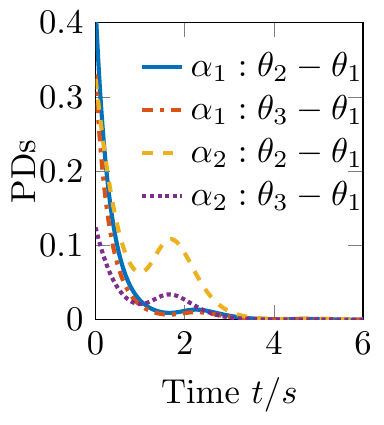}
		\label{Fig:4}}
	\subfigure[]{
		\includegraphics[scale=0.62]{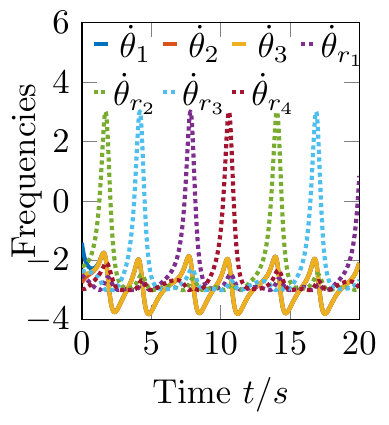}
		\label{Fig:5}}
	\subfigure[]{\begin{tikzpicture} [->,>=stealth',shorten >=1pt,auto,node distance=0.5cm,
		main node/.style={circle,fill=blue!5,draw,minimum size=0.3cm,inner sep=0pt]},
		red node/.style={circle,fill=red!30,draw,minimum size=0.3cm,inner sep=0pt]},
		white node/.style={circle,fill=white,draw,minimum size=0.3cm,inner sep=0pt]}]
		
		\node[red node]          (0) at (-0.55,0)                      {\scriptsize $r_2$};
		\node[red node]          (7) at (-0.55,0.4)                      {\scriptsize $r_1$};
		\node[red node]          (8) at (-0.55,-0.4)                      {\scriptsize $r_3$};
		
		\node[white node]          (4) at (0.45,0.4)                      {\scriptsize $4$};
		\node[white node]          (10) at (0.45,-0.4)                      {\scriptsize $8$};
		\node[white node]          (3) at (0.45,1)                      {\scriptsize $3$};
		\node[white node]          (5) at (0.45,-1)                      {\scriptsize $5$};
		
		\node[white node]          (6) at (-0.55,1)                      {\scriptsize $6$};
		\node[white node]          (9) at (-0.55,-1)                      {\scriptsize $7$};
		\node          (15) at (-0.55,-1.2)                      {};
		
		\node[main node]          (2) at (-1.2,-0.3)         {\scriptsize $2$};		
		\node[main node]          (1) at (-1.2,0.3)          {\scriptsize $1$};
		

		\tikzset{direct/.style={-,line width=1pt}}
		\tikzset{DasDirect/.style={-,dashed,red,line width=1pt}}
		\path (0)     edge[direct]     node   {} (1) 
		(0)     edge[direct]     node   {} (2)
		(8)     edge[DasDirect]     node   {} (1)
		(8)     edge[direct]     node   {} (9) 
		(8)     edge[direct]     node   {} (5) 
		(5)     edge[direct]     node   {} (9)
		(5)     edge[direct]     node   {} (10)
		(4)     edge[direct]     node   {} (10) 
		(8)     edge[DasDirect]     node   {} (2) 
		(7)     edge[DasDirect]     node   {} (1)
		(3)     edge[direct]     node   {} (4)
		(7)     edge[DasDirect]     node   {} (2)
		(7)     edge[direct]     node   {} (6)
		(0)     edge[direct]     node   {} (3)
		(3)     edge[direct]     node   {} (6);
		\end{tikzpicture} \label{general}}
	\subfigure[]{
		\includegraphics[scale=0.62]{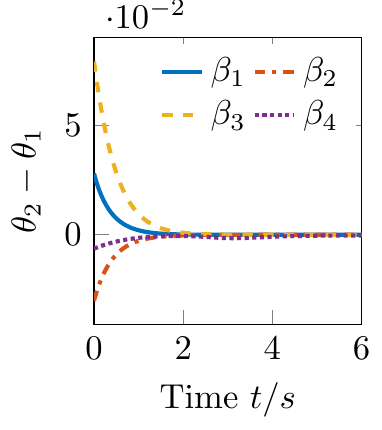}
		\label{general:result}}	
	\subfigure[]{
		\includegraphics[scale=0.62]{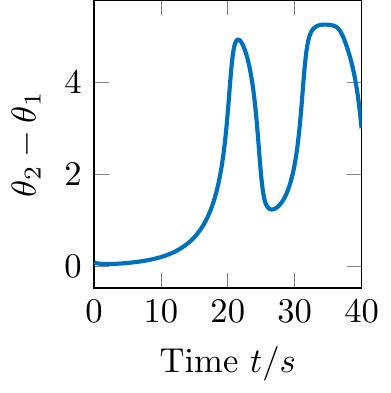}
		\label{general:result2}}
	
	\caption{(a) trajectories of the phase differences (PDs) when $n=m=3$; (b) trajectories of PDs when $n=3,m=4$ and $\alpha=\alpha_1, \alpha_2$; (c) trajectories of the frequencies when $n=3,m=4$ and $\alpha=\alpha_1$;  (d) a network with a bipartite component ($1,2$ mediated by $r_1,r_2,r_3$); (e) trajectories of $\theta_2-\theta_1$ for $\alpha=\beta_1,\beta_2,\beta_3$ with red dashed edges in (d); (f) trajectories of $\theta_2-\theta_1$ for $\alpha=0.9$ without red dashed edges in (d).}
\end{figure}
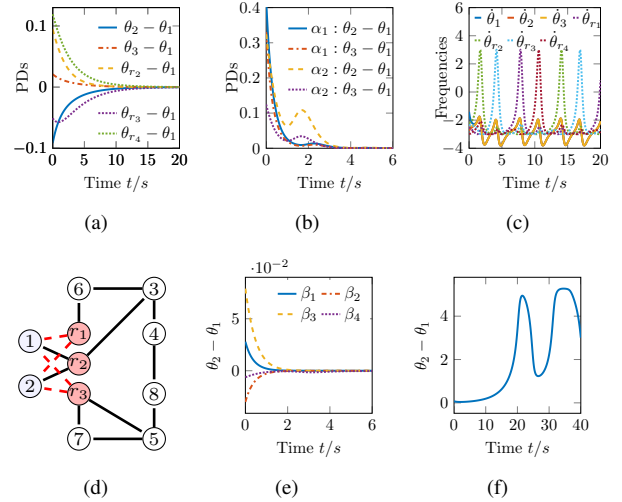

	\section{Simulation Results}\label{simulation}
	In this section, we present a set of illustrative simulations that go beyond our theoretical results, and provide some interesting observations.

	First, we investigate the situation when $m\ge n$.	Phase differences in the case of $m=n=3$ are plotted in Fig. \ref{Fig:3}. It appears that \textit{complete} (not just remote) synchronization can occur for any $\alpha\in(0,\pi/2)$, since it is observed for a very large $\alpha$, i.e., $\pi/2-0.1$.  When $n=3$ and $m=4$, it can be observed from Fig. \ref{Fig:4} that phase synchronization of the mediated oscillators always takes place even when the phase shift $\alpha$ is large ($\alpha_1=1.35,\alpha_2=1.4$). However, the frequencies of the mediated oscillators' converge to a dynamically changing value that is quite distinct from those of the  mediators (see Fig. \ref{Fig:5}). This implies that phase-unlocked remote synchronization has occurred. Moreover, the frequencies of the mediators stay distinct from one another, which implies that their phases also remain distinct. This phenomenon is known as a Chimera state \cite{PMJ-ADM:15}, since synchronization and desynchronization coexist in the same network. Interestingly, simulation results confirm the occurrence of remote synchronization for any $\alpha$. We conjecture that the mediators' quantitative advantage creates a powerful structure, which can eliminate the effect of any phase shift or time delay, making remote synchronization always stable.

	Clearly, complete bipartite networks are a special class of networks. Real networks, such as brain networks, are certainly not bipartite. Yet, bipartite components can be found in complex networks including brain networks, and reasonably might play a role in enforcing synchronization. Then, we consider a network with a bipartite subgraph in Fig. \ref{general}. It is shown in Fig. \ref{general:result} that the oscillators $1$ and $2$, mediated by $r_1$, $r_2$ and $r_3$, gradually become synchronized for a wide range of phase shift $\alpha$ (the cases when $\alpha$ equals $\beta_1=0.6,\beta_2=0.8,\beta_3=1,$ and $\beta_4=1.2$ are plotted). However, if we reduce the number of mediators by removing the red dashed edges in Fig. \ref{general}, we find that remote synchronization cannot appear anymore even when the phase shift $\alpha$ is as small as $0.9$. From the simulations, we confirm that a bipartite subgraph in a network plays an important role in ensuring stable remote synchronization. Moreover,  more mediators make remote synchronization more robust against phase shifts (or time delays), as suggested by our theoretical findings for bipartite networks.

\section{Conclusion}\label{conclusion}
Cortical regions without apparent neuronal links exhibit synchronized behaviors, and the common neighbors that they share seem to play crucial roles in giving rise to this phenomenon. Motivated by these empirical observations, we have analytically studied mediated remote synchronization of Kuramoto-Sakaguchi oscillators coupled by bipartite networks. A larger number of mediators has been shown to make remote synchronization more robust to phase shifts or time delays. Simulation results confirm that this finding also applies to more complex networks with bipartite components. Moreover, remote synchronization seems to be stable for any phase shift if there are more mediators than the mediated oscillators in a bipartite network. This aspect is left as the subject of future investigation.

\appendix

\begin{lemma}\label{lemma:2}
	Let $c(\alpha)$ be defined in \eqref{expre:c(a)}, then 
	$
	\cos(c(\alpha)+\alpha)>0
	$
	for any $m_1$ and any $\alpha\in(0,\pi/2)$.
\end{lemma}
\begin{IEEEproof}
	There holds that $\cos(c(\alpha)+\alpha)=\cos c(\alpha)\cos \alpha-\sin c(\alpha)\sin \alpha$. Substituting \eqref{cos:sin} into the right-hand side of this inequality we can compute $\cos(c(\alpha)+\alpha)=\frac{1}{S}( n-m_1+2m_1\cos ^2\alpha +m_2\cos 2\alpha )=\frac{1}{S}(n-m+2m \cos^2\alpha)$,	where the last equality has used the double-angle formula $\cos 2\alpha=2\cos^ 2\alpha -1$ and the equality $m_1+m_2=m$. Then, $\cos(c(\alpha)+\alpha)>0$ for any $\alpha$ and $m_1$ since $m<n$ by hypothesis.
\end{IEEEproof}

\begin{proposition}\label{Poly:>0}
	Let $a=\cos (c(\alpha)+\alpha)$, $b=\cos(c(\alpha)-\alpha)$, and $c=-\cos(c(\alpha)+3\alpha)$, where $c(\alpha)$ is given by \eqref{expre:c(a)}. Given $m\ge 2$, suppose $m_1=m-1$,  then $(m-1)b+na-c>0$ for any $\alpha\in(0,\pi/2)$.
\end{proposition}


\begin{IEEEproof}
	Substituting \eqref{expre:c(a)} into $(m-1)b+na-c$ and after some algebra one can obtain $S\big((m-1)b+na-c\big)=(n-m+1)\big(n-m-1+2(m+1)\cos^2 \alpha\big)\big)+2(m-1)\big(8 \cos^4 \alpha+(n+m-7)\cos^2\alpha\big)+1.$
	If $m= 2$, then $n\ge 3 $ by hypothesis, and subsequently, $n-m+1\ge2$, $n-m-1\ge 0$, and $n+m-7\ge -2$, which means that 
	$
	S\big((m-1)b+na-c\big)\ge 16 \cos^4 \alpha +8\cos^2\alpha+1>0.
	$
	If $m\ge 3$, then $n\ge 4 $, which subsequently means that $n-m+1\ge 2$, $n-m-1\ge 0$, and $n+m-7\ge 0$. Consequently,
	$
	S\big((m-1)b+na-c\big)\ge 4(m+1)\cos^2\alpha+ 16(m-1) \cos^4 \alpha+1>0.
	$
	Since $S>0$, it follows naturally that $(m-1)b+na-c>0$.
\end{IEEEproof}

\bibliographystyle{IEEEtran}
\bibliography{\alias,\New,\Main,\FP}

\end{document}